\documentclass[useAMS,usenatbib,letters]{mn2e}
\usepackage{aas_macros,graphicx,times,multirow,amsmath}
\usepackage[]{color,hyperref}

\title[The variable X-ray emission  of \src]{The variable X-ray emission of    \src   }
\author[S. Mereghetti et al.]{S.~Mereghetti,$^{1}$\thanks{E-mail: \href{mailto:sandro@iasf-milano.inaf.it}{sandro@iasf-milano.inaf.it}}
 A.~Tiengo,$^{2,1,3}$ P.~Esposito$^{1}$, R.~Turolla$^{4,5}$
\smallskip\\
$^1$INAF -- Istituto di Astrofisica Spaziale e Fisica Cosmica - Milano, via E. Bassini 15, I-20133 Milano, Italy\\
$^2$IUSS -- Istituto Universitario di Studi Superiori, piazza della Vittoria 15, I-27100 Pavia, Italy\\
$^3$INFN -- Istituto Nazionale di Fisica Nucleare - Pavia, via Bassi 6, I-27100 Pavia, Italy\\
$^4$Dipartimento di Fisica e Astronomia, Universit\`a di Padova,
Via Marzolo 8, I-35131 Padova, Italy\\
$^5$Mullard Space Science Laboratory, University College London, Holmbury St. Mary, Dorking, Surrey, RH5 6NT, UK
}
\date{Accepted \ldots. Received \ldots; in
original form \ldots} \pagerange{\pageref{firstpage}--\pageref{lastpage}} \pubyear{2011}

\def\LaTeX{L\kern-.36em\raise.3ex\hbox{a}\kern-.15em
    T\kern-.1667em\lower.7ex\hbox{E}\kern-.125emX}

\def\xmm {\emph{XMM-Newton}}

\def\src {PSR\,B0943+10}
\def\psr {PSR\,B0943+10}

\def\pdot {\dot P}

\begin{document}

\label{firstpage}
\maketitle
\begin{abstract}
 The old pulsar \src\ ($P$=1.1 s, characteristic age $\tau$=5 Myr)  is the best example of mode-switching radio pulsar.
 Its radio emission alternates between a highly organised state  with regular drifting subpulses (B mode)
 and a   chaotic emission pattern (Q mode). We present the results of  \xmm\ observations showing that
 the X-ray properties of \psr\ depend on its radio state \citep{her13}.
 During the radio fainter state (Q mode) the X-ray flux is more than a factor two larger than during the
 B-mode and X-ray pulsations with $\sim$50\% pulsed fraction are detected.
 The X-ray emission of \psr\   in the B-mode is well described by thermal emission with   blackbody  temperature $kT$=0.26 keV
 coming from a small hot spot with luminosity of 7$\times10^{28}$ erg s$^{-1}$, in good agreement with the prediction of the partially screened
 gap model, which also explains the properties of the radio emission in this mode. We derived an upper limit of 46\%
 on the X-ray pulsed fraction in the B-mode, consistent with the geometry and viewing angle of \psr\ inferred from the
 radio data.
 The higher flux observed during the Q-mode is consistent with the appearance of an additional component with
 a power-law spectrum with photon index 2.2.
 We interpret it as pulsed non-thermal X-rays produced  in the star magnetosphere.
 A small change in the beaming pattern or in the efficiency of  acceleration of the particles responsible for the
 non-thermal emission can  explain the reduced flux of this component during the radio B-mode.
\end{abstract}
\begin{keywords}
pulsars: general -- stars: neutron -- X-rays: individual: \src.
\end{keywords}

\section{Introduction}

Rotation-powered neutron stars are observed as sources of thermal and non-thermal X-rays, whose
relative intensity and   properties mainly depend on the star age, spin period and magnetic field  \citep{bec09}.
Non-thermal emission is produced by particles accelerated in the star magnetosphere at the
expense of rotational energy.
The thermal X-rays observed in middle-aged pulsars are due to the dissipation of the internal heat, while
older neutron stars show thermal emission from hot polar caps heated by relativistic particles
accelerated in the magnetosphere and flowing toward the star surface.

Several decades of X-ray  observations  led to the widespread notion that,
except for the periodic variability due to the star spin,   rotation-powered neutron stars  are steady X-ray emitters.
The only  example of X-ray variability (a few short bursts associated to
a flux enhancement lasting  about one month)  observed in an allegedly rotation-powered pulsar has been attributed to
the release of magnetic energy (PSR J1846--0258, \citealt{gav08}).
In fact, X-ray variability on time scales from few milliseconds to months  is  a characteristic of
isolated neutron stars whose emission is powered by magnetic energy rather than by the loss of rotational energy,
the so called magnetars \citep{mer08}.
On the other hand, it is well known that pulsars can
exhibit significant variability on different time scales at radio wavelengths \citep{kea13}.
In particular many pulsars alternate between two  states which differ in the shape of their
mean radio pulse profiles (see, e.g. \citealt{lor12}, and references therein).

The recent discovery of  variations by a factor $\sim$2  in the
X-ray luminosity of  \src , correlated with changes in its mode of
radio emission \citep{her13}, is at variance with the picture
outlined above and has come as a major surprise. \src\ is a
relatively old radio pulsar with spin period $P=1.1$ s and period
derivative $\pdot$=3.5$\times$10$^{-15}$ s s$^{-1}$. These timing
parameters imply a characteristic age  $\tau$=$P/(2\pdot$) = 5
Myr, a dipolar magnetic field $B$=4$\times$10$^{12}$ G, and a
rotational energy loss rate $\dot{E}_{rot}$ = 10$^{32}$ erg
s$^{-1}$. In the radio band \src\ alternates between two different
states: when the pulsar is in the so called B  (burst-like or bright)
mode, the radio emission displays a regular pattern of drifting
subpulses, while it is chaotic, and on average fainter, when the
pulsar is in the Q (quiescent) mode \citep{sul84}.

X-rays from \src\ were first detected in two observations carried out in 2003 with \xmm\ \citep*{zha05}, but due to the
source faintness and short exposure times it was not possible  to precisely characterize the
spectrum and detect X-ray pulsations.
Further \xmm\ pointings of \src\ were carried out in 2011, with simultaneous radio coverage at 320 MHz
with the Giant Metrewave Radio Telescope (GMRT) and at 140 MHz with the Low Frequency Array (LOFAR).
These observations showed that the pulsar X-ray emission is
brighter during the radio Q mode and fainter during the B mode.
X-ray pulsations were detected only when the flux was higher.
According to the spectral  analysis of \citet{her13},   the X-ray data can be interpreted assuming that
when the pulsar is in the B-mode it emits only a non-thermal, unpulsed  power-law component and the higher flux
in the Q-mode is due the additional presence of a 100\%-pulsed thermal component, well described by a blackbody model.

Here we present an independent reanalysis of the 2011 \xmm\ observations which confirms the main results of \citet{her13}, but
shows that other models not considered by these authors and implying different,   possibly  more natural, interpretations
are also consistent with the  data.
The distance of \src , based on its dispersion measure and the Cordes \& Lazio (2002)
electron density model, is 630$\pm$100 pc. In the following we assume this value for all the distance-dependent quantities.

\begin{table*}
\begin{minipage}{11.5cm}
\centering \caption{Log of the \xmm\ observations}
\label{obs-log}
\begin{tabular}{@{}cccrrr}
\hline
 Obs.ID & \multicolumn{2}{c}{Start/end time (UT)}   & \multicolumn{3}{c}{Net exposure time (ks)} \\
        & \multicolumn{2}{c}{(YYYY-MM-DD hh-mm-ss)} &          pn    &     MOS1  & MOS2  \\
\hline
 0671540201  & 2011-11-04 01:16:13  & 2011-11-04 07:16:23  & 13.4 & 21.0 &  21.0 \\
 0671540301  & 2011-11-06 01:18:32  & 2011-11-06 08:28:53  & 18.2 & 23.9 &  24.4 \\
 0671540401  & 2011-11-21 23:54:31  & 2011-11-22 06:44:42  & 15.8 & 19.5 &  21.9 \\
 0671540501  & 2011-11-27 23:52:59  & 2011-11-28 06:04:42  &  7.6 & 20.5 &  20.7 \\
 0671540601  & 2011-12-01 23:39:03  & 2011-12-02 05:52:36  &  9.1 & 15.7  & 16.1 \\
\hline
\end{tabular}
\end{minipage}
\end{table*}

\begin{table*}
\begin{minipage}{11.5cm}
\centering \caption{Results of the spectral fits}
\label{spectra}
\begin{tabular}{@{}lccccccc}
\hline
  Parameter  & \multicolumn{4}{c}{Q mode } & &\multicolumn{2}{c}{B mode } \\
             &  PL & BB & PL+BB &   & &PL & BB  \\
\hline
N$_H$  (10$^{20}$ cm$^{-2}$)               & 14$_{-5}^{+6}$     & $<$1    &  $<$6           &              && 33$_{-15}^{+23}$  &  $<$20    \\
Photon index                                 & 3.1$_{-0.3}^{+0.4}$ &  -      & 2.6$_{-0.4}^{+1.1}$ &             &&  4.1$_{-1.0}^{+1.5}$ & -     \\
K$_{PL}^{(a)}$ (10$^{-6}$ ph cm$^{-2}$ s$^{-1}$ keV$^{-1}$) & 7.0$_{-1.3}^{+1.8}$ &  -   & 1.6$_{-0.8}^{+1.2}$ &  &&  5.7$_{-2.3}^{+6.2}$  & - \\
kT$_{BB}$ (keV)                               &  -              &0.26$\pm$0.02 & 0.31$\pm$0.04  &   &&  -   & 0.27$_{-0.02}^{+0.03}$     \\
F$_{BB}^{(b)}$  (10$^{-15}$ erg cm$^{-2}$ s$^{-1}$)    &  -    & 13.5$\pm$0.6     & 8.7$_{-2.3}^{+2.8}$ & && -& 5.8$\pm$0.6    \\
R$_{BB}^{(c)}$  (m)                              &  -            & 30.7$_{-3.3}^{+3.7}$     & 18.5$_{-4.5}^{+5.1}$ & && -& 19.5$_{-3.5}^{+4.0}$    \\
$\chi_{\nu}^2$/dof                               & 1.00 / 20           & 1.34 / 20     &  0.70 / 18             &          && 0.39 / 7 & 0.47 / 7  \\
Null hypothesis probability              & 0.46                & 0.14    &  0.81         &                     &&  0.91    &    0.86  \\
 \hline
\end{tabular}
\begin{list}{}{}
\item Errors at 1$\sigma$ for one interesting parameter.
\item[$^{a}$]
 Flux of the power-law at 1 keV.
 \item[$^{b}$]
 Bolometric blackbody flux.
 \item[$^{c}$]
 Blackbody radius for d=630 pc.
\end{list}
\end{minipage}
\end{table*}

\section{Observations and data analysis}

\subsection{Observations and data selection}

Six observations of \src\ were performed with \xmm\ between 2011 November 4 and December 4.
We used the data obtained with the EPIC instrument which consists of one pn \citep{str01}
and two  MOS CCD cameras \citep{tur01}. In all the observations the
pn camera was operated in full frame mode (73 ms resolution), and  both MOS
cameras in small window mode (0.3 s resolution). The thin optical blocking filters were used.

We checked the data for the presence of time intervals with high background due to soft proton flares
and removed them,   resulting in the net exposure times given in Table   \ref{obs-log}.
We discarded the last observation which was completely affected by high background.

The extraction of the counts for the
spectral and timing analysis  was done from a circular region of radius 15$''$ centered on the
pulsar position, both for the pn and the two MOS.
We considered only photons in the 0.2-12 keV energy range and with  patterns
corresponding to mono- and bi-pixel events
(i.e. $\leq$4 for pn and $\leq$12 for MOS).
The background spectra were extracted from  source-free regions in the same CCD chips used for the target.
Different spectra were extracted for the  two radio states of \src ,
based on the time intervals   given in \citet{her13}. These spectra   were then summed to produce
a total spectrum for each camera  and source state. The corresponding net exposure times are similar for the Q and B mode:
32.9  and 31.3 ks respectively for the pn,
51.7 and 48.9  ks for the MOS1,
55.4 and 48.6 ks for the MOS2.

\subsection{Timing and spectral analysis}

All the spectra were rebinned  to have a minimum of 20  counts in each energy channel after background subtraction and
were fitted using  the XSPEC package V12.7. For the interstellar absorption we used the \textsc{phabs} model.
The spectra of the three cameras were fitted simultaneously.

\begin{figure}
\includegraphics[width=60mm,angle=-90]{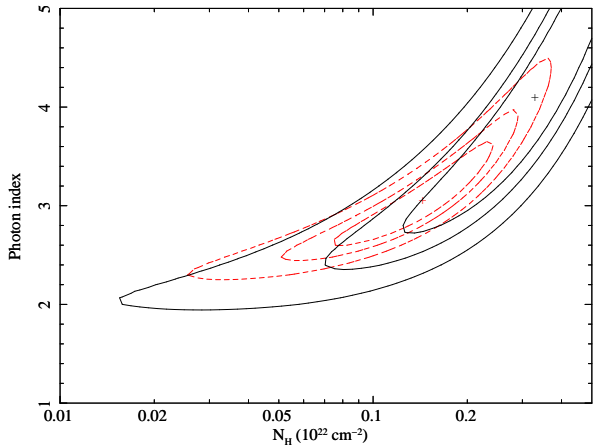}
\caption{Error regions of absorption and photon index  (68\%, 90\% and 99\% c.l. for two interesting parameters) for the power-law model  fits
of the B-mode (black solid lines) and Q-mode (red dashed lines) spectra.}
\label{figQandB_PLc}
\end{figure}

\begin{figure}
\includegraphics[width=60mm,angle=-90]{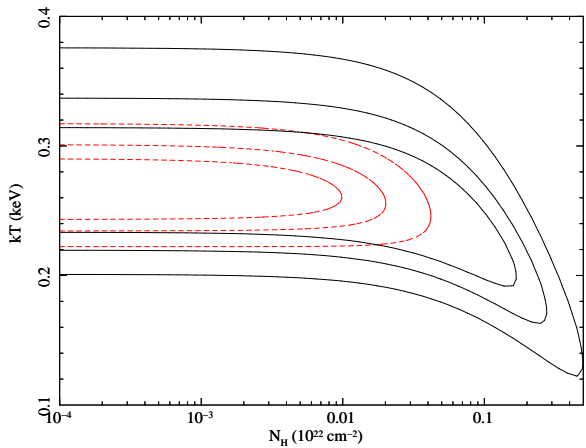}
\caption{Error regions of absorption and temperature  (68\%, 90\%
and 99\% c.l. for two interesting parameters) for the blackbody model  fits of the B-mode (black
solid lines) and Q-mode (red dashed lines) spectra.}
\label{figQandB_BBc}
\end{figure}

Good fits were obtained with either a power law or a blackbody model for the  spectra of both modes.
We did not find statistically significant evidence for  spectral variations between the Q and B mode, as it can be seen
in  Figs. \ref{figQandB_PLc} and \ref{figQandB_BBc}, where the confidence regions of the fit parameters are plotted.
In fact, the best fit Q-mode models provide an adequate fit also to the B-mode spectra by changing only their normalizations.
The flux in the B-mode is about 42\% of that in the Q-mode.
Although single component models are adequate to describe the data, we
fitted  the Q-mode spectra also with a power law plus blackbody.
All the spectral results are summarized in  Table  \ref{spectra}.

 \begin{figure}
 \includegraphics[width=120mm,angle=-90]{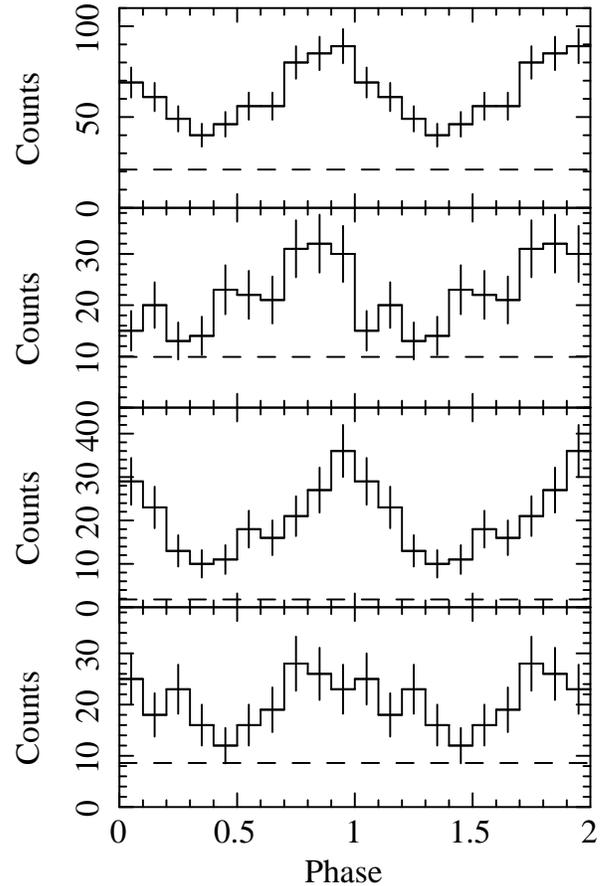}
\caption{Folded light curves (pn+MOS) of \src\ in the Q mode  in the energy ranges 0.15-12 keV,
0.15-0.6 keV,
0.6-1.3 keV,
1.3-12. keV  (top to bottom). The dashed lines indicate the estimated background level.}
 \label{figLC_QQQ}
 \end{figure}

 \begin{figure}
 \includegraphics[width=60mm,angle=-90]{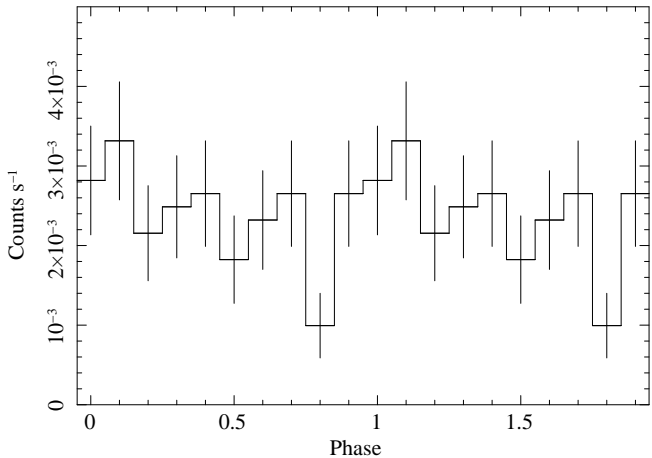}
\caption{Folded pn light curve  of \src\ in the B mode  (0.15-12 keV).}
 \label{figLC_BBB}
 \end{figure}

After converting the time of arrivals to the Solar system
barycenter, we folded at the spin period of \psr\ the pn and MOS
counts obtained during  the Q-mode. We used the timing parameters
of \citet{her13} ($\nu$=0.910989538329    Hz, d$\nu$/dt =
--2.94219$\times10^{-15}$  Hz s$^{-1}$, d$^2\nu$/dt$^2$ =
--1.39$\times10^{-25}$ Hz s$^{-2}$, epoch = 54226 MJD). The
resulting pulse profile in the 0.15--12 keV  energy range is shown
in the top panel of   Fig. \ref{figLC_QQQ}. The pulse
profiles in the soft (0.15--0.6 keV),  medium (0.6--1.3 keV) and
high (1.3--12 keV) energy ranges are also reported (lower panels).
The three energy intervals were chosen to have approximately the
same number of counts (about 200) in each band. The horizontal
lines indicate the estimated background levels.

To derive the significance of the pulse detection, we computed the
$Z^2_1$ test statistics \citep{buc83} and obtained values
corresponding to single trial chance probabilities of
2.5$\times$10$^{-8}$ ($Z^2_1$=35.0), 2.2$\times$10$^{-3}$ ($Z^2_1$=12.2),  7.1$\times$10$^{-6}$ ($Z^2_1$=23.7),
and  2.2$\times$10$^{-2}$ ($Z^2_1$=7.6) in the total, soft, medium, and hard
band, respectively. A fit with a constant plus a sinusoid gives a
pulsed fraction in the total band of  (50$\pm$6)\%. The pulsed
fractions in the other bands are  (66$\pm$13)\% soft, (56$\pm$8)\%
medium, and  (53$\pm$13)\% hard.

We analyzed in a similar way the counts extracted when the pulsar
was in  radio B-mode (350   counts in the total energy range). No
significant modulation was detected in this case (values of Z$^2_1$$<$2.54, corresponding to
probabilities larger than 0.28, were obtained in the different
energy ranges). Through Monte Carlo simulations and assuming a
sinusoidal modulation, we derived 3$\sigma$ upper limits on the
pulsed fraction during the B-mode of 56\% and 46\% in the medium
and total energy ranges, respectively. The folded light curve in the B-mode is
shown in Fig. \ref{figLC_BBB}.

For the phase-resolved spectroscopy of the Q-mode data we extracted the pn spectra of two non-overlapping phase
intervals of equal width ($\Delta \phi$=0.5), centered at the
maximum and minimum of the pulse profile in the total energy range.

\begin{figure}
 \includegraphics[width=85mm,angle=-90]{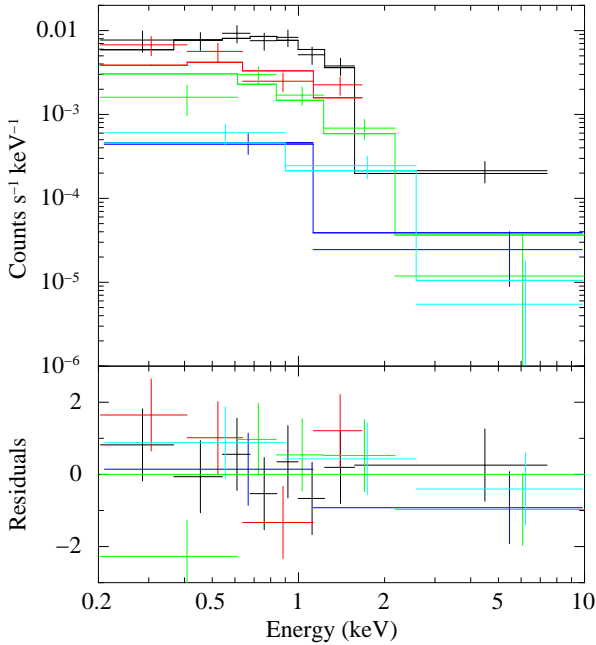}
\caption{Fit with a constant power law plus a variable blackbody
to the B-mode spectrum (green, blue and cyan for pn, MOS1 and MOS2, respectively)
and to the pn Q-mode spectra of the pulse maximum (black) and minimum (red).  }
 \label{B_Qpul_Qunp_constPL_pulsedBB}
 \end{figure}

We make the working hypothesis that the difference in the X-ray
flux between the  two radio states of \psr\ is due only to the
presence during the Q-mode of an additional pulsed component,
which is absent (or too faint to be detected) during the B-mode.
Different possibilities are explored by fitting simultaneously the
two phase-resolved Q-mode spectra (pn only) and the B-mode
spectrum (pn+MOS). We first consider the case of a power-law which
remains steady plus a pulsed blackbody: this is done by imposing
common parameters for the power-law component in the three spectra
and a blackbody normalization fixed to 0 for the B-mode spectrum.
In this way  a good fit is found with a power-law photon index of
2.3$\pm$0.4 and a $kT=0.26\pm$0.04  keV blackbody from an emitting area which
varies by a factor of $\sim$4 between the pulse minimum and maximum.
The best   fit is shown in Fig. \ref{B_Qpul_Qunp_constPL_pulsedBB}.

\begin{figure}
 \includegraphics[width=85mm,angle=-90]{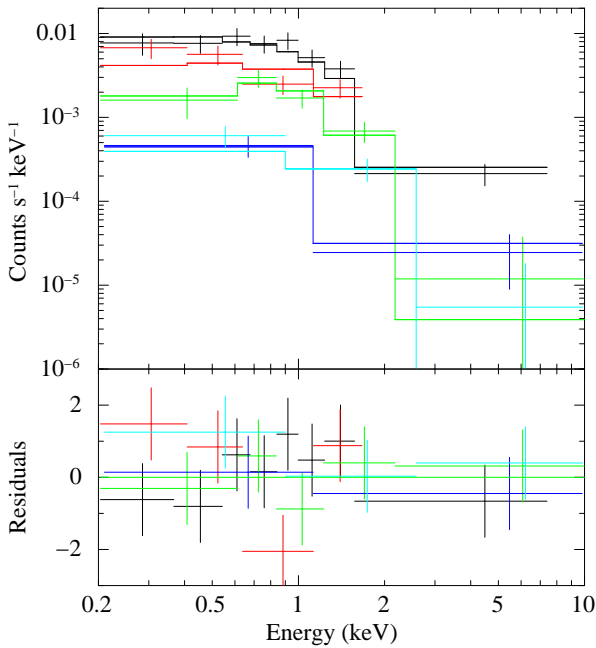}
\caption{Fit with a constant blackbody plus a variable power law
to the B-mode spectrum (green, blue and cyan for pn, MOS1 and MOS2, respectively)
and to the pn Q-mode spectra of the pulse maximum (black) and minimum (red).  }
 \label{B_Qpul_Qunp_constBB_pulsedPL}
 \end{figure}

In the same way we explored the alternative possibility of a steady blackbody component plus  a
variable power-law responsible for the Q-mode pulsations. A slightly better fit was obtained in this case,
with similar values of temperature and photon index
(Fig. \ref{B_Qpul_Qunp_constBB_pulsedPL}).
All the best-fit results are summarized in  Table \ref{spectrap2}.  A comparison of the two models is shown
in Fig. \ref{models}.

\begin{table}
\begin{minipage}{11.5cm}
\caption{Results of simultaneous  fits to the B- and Q-mode }
\label{spectrap2}
\begin{tabular}{@{}lcc}
\hline
    Parameter                   & Constant PL          & Constant BB     \\
                                & and pulsed BB        & and pulsed PL  \\
\hline
N$_H$  (10$^{20}$ cm$^{-2}$)                             & 4.3$_{-3.3}^{+4.7}$          &  2.4$_{-2.4}^{+3.5}$          \\
Photon index                                             & 2.3$_{-0.3}^{+0.4}$           & 2.2$\pm$0.3                 \\
K$_{PL}^{(a)}$ (10$^{-6}$ ph cm$^{-2}$ s$^{-1}$ keV$^{-1}$)  & 2.0$_{-0.4}^{+0.6}$      &  1.4$\pm$0.5   /    4.0$_{-0.8}^{+1.1}$            \\
kT$_{BB}$ (keV)                                          & 0.26$\pm$0.04                 &    0.26$\pm$0.03                \\
F$_{BB}^{(b)}$  (10$^{-15}$ erg cm$^{-2}$ s$^{-1}$)      & 4.2$\pm$1.6 / 16.6$_{-2.7}^{+3.5}$   &    6.3$_{-0.8}^{+1.0}$      \\
R$_{BB}^{(c)}$  (m)                   &  17$\pm$7 / 35$\pm$11  &    21$\pm$6          \\
$\chi_{\nu}^2$/dof                                             &  1.16 / 16                   & 0.99 / 16            \\
Null hypothesis probability                              &  0.29                  &   0.47      \\
\hline
\end{tabular}
\begin{list}{}{}
\item Errors at 1$\sigma$ for one interesting parameter.
\item[$^{a}$]
 Flux of the power-law at 1 keV.
 \item[$^{b}$]
 Bolometric blackbody flux.
 \item[$^{c}$]
 Blackbody radius for d=630 pc.
\end{list}
\end{minipage}
\end{table}

\section{Discussion}

Our analysis of the 2011 \xmm\ observations   confirms that the
X-ray flux of \psr\ changes on short timescales in strict
correlation with the two modes of radio emission which
characterize this pulsar \citep{her13}.
The observed  0.2--10 keV flux is about 5$\times10^{-15}$ erg cm$^{-2}$ s$^{-1}$ during the   B-mode,
in which the radio emission shows a highly organized structure with drifting subpulses.
When the pulsar switches  to  the radio Q-mode, its  X-ray flux increases by a factor $\sim$2.5.
X-ray pulsations with a nearly sinusoidal profile and a pulsed fraction of $\sim$50\%
are clearly present when the X-ray flux is high.
On the other hand, we could not detect the pulsations in the sum of all the data collected during the  radio B-mode.
Due to the X-ray faintness of the source in this state, the derived upper limits
on the pulsed fraction
are not particularly constraining.

The spectrum of \psr\ during the B-mode is fit equally well   by
either a power law or a blackbody. The power-law model yields a
rather steep spectrum, photon index of 4.1,  and an interstellar  absorption larger than
that expected based on the total Galactic column density in this
direction, $\sim2.3\times10^{20}$ cm$^{-2}$ \citep{kal05}.
Although the involved uncertainties do not allow us to reject the
power-law model on these bases, we favour the blackbody
interpretation, which yields a temperature and emitting area consistent with the less constrained values derived with the 2003 data  \citep{zha05}.
Note that the  radio state of the pulsar during the 2003 observations is not known, due to the lack of simultaneous radio coverage.

Detailed modeling  of the radio data indicates  that
\src\ is a nearly aligned rotator (the angle between the rotation and
magnetic axis is of $\sim$15$^{\circ}$) seen nearly pole-on \citep{des01}.
This implies that one of the magnetic polar cap regions is always visible
and the pulsar rotation causes only a minor modulation of the observed  flux.
We computed the light curve produced by a single rotating hot spot on the star
surface under the conditions expected in \src, including general
relativistic effects, and found a pulsed fraction $\la 2$\%,
consistent with the lack of strong pulsations in B-mode.

\begin{figure}
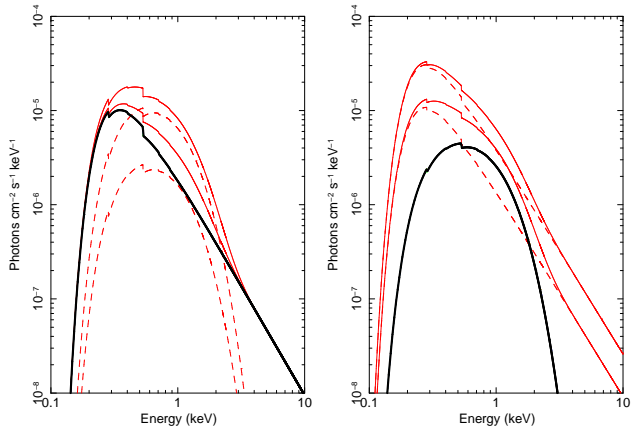

\resizebox{\hsize}{!}{\includegraphics[angle=-90]{B_Qpul_Qunp_constPL_pulsedBB_model.ps}
\includegraphics[angle=-90]{B_Qpul_Qunp_constBB_pulsedPL_model.ps}}
\caption{ Best-fit  power-law plus blackbody model. The thick black lines indicate the  constant components which fit the B-mode spectra:
either a power-law (left panel) or a blackbody (right panel). The additional pulsed components present only in the Q-mode are indicated by the
red dashed lines: either a blackbody (left panel) or a power-law (right panel). The red solid lines are the total spectra of the Q-mode at the
maximum and minimum of the pulsations. }
\label{models}
\end{figure}

Although with the statistical quality of the current data there is
no compelling evidence for differences between the X-ray spectra
of the two states, it is interesting to explore such a
possibility. Since most well studied pulsars require a combination
of thermal and non-thermal components in their X-ray spectra, we assumed
that the same applies to \psr ,
and checked whether  the
higher, and pulsed,  X-ray flux during the Q-mode can be explained
by the appearance of a new spectral component in addition to that
of the B-mode, which is assumed to be always present with the same
properties in both modes.
The simplest   picture is  that of a  blackbody plus
power-law spectrum in which  either the blackbody or the power-law
is the ''additional pulsed component'' (see Fig. \ref{models}), As shown above, both cases are consistent with the data.

The first case has been considered by \citet{her13}. In this
scenario it is easy to account for the constant power-law
component, which could be well explained by non-thermal processes
in the star magnetosphere or in a pulsar wind nebula. On the other
hand it seems difficult to produce a large modulation of the thermal
component at the spin period.   As discussed above, thermal,
isotropic emission from a small heated region on the star surface
results in a pulsed fraction more that 20 times lower than the
observed one,
given that \src\ is a nearly aligned
rotator seen close to the rotation axis. Furthermore, some
mechanism is required to switch-off, or strongly reduce, this
component during the Q-state. According to \citet{her13}, a 100\%
modulation of the blackbody component can be
produced by absorption/scattering in the magnetosphere, assuming
that its properties vary with the pulsar rotation phase. The same
process could operate at all rotational phases in the B-mode, e.g.
because the region of closed field lines expands and completely
blocks our view of the polar cap, explaining the non detection of
the thermal component. Even leaving aside what actually produces
the changes in the pulsar magnetosphere, this picture would imply
the presence of a large enough charge density in the closed
magnetosphere. Resonant cyclotron scattering on
electrons/positrons (by far the most efficient process) requires a
particle density $\approx 10^{14}$--$10^{15}\, \mathrm{cm}^{-3}$
to make the region optically thick. This is comparable to the
charge density in the twisted magnetosphere of magnetars, in which
substantial deviations from the dipole field geometry occur
because of the huge magnetic stresses exerted on the crust by the
internal field
\cite*[e.g.][]{tho02}.
Would this occur, \src\ ought to be akin (at least in some
respects) to the soft $\gamma$-repeaters and anomalous X-ray
pulsars, which have, however, a largely different phenomenology
at X-ray energies.
Alternatively, one may assume that
the back-flow of particles which heat the pole immediately
switches off at the transition between the Q- and B- mode,
although this leaves the pulsed fraction issue open.

\begin{figure}
 \includegraphics[width=90mm,angle=0]{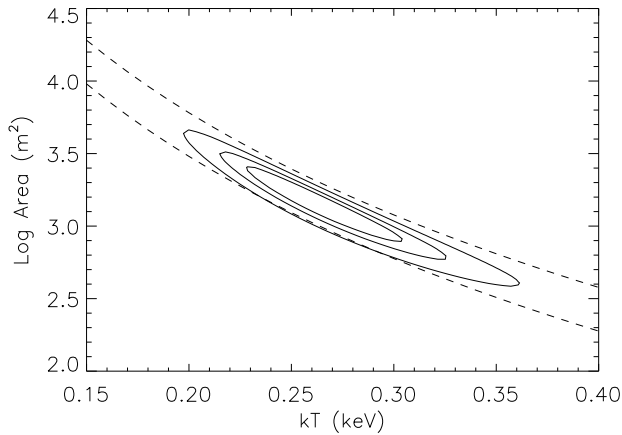}
\caption{Error regions of emission area and temperature for the steady blackbody component (68\%, 90\% and 99\% c.l. for two interesting parameters).
The dashed lines correspond to bolometric luminosities of   5$\times10^{28}$  and  $10^{29}$ erg s$^{-1}$.  }
 \label{fig_area_kT}
 \end{figure}

Our analysis clearly shows that also the second
of the two possibilities discussed above, i.e. a pulsed power-law
plus a constant blackbody, is fully consistent with the data (and
even slightly preferred  in terms of $\chi^2$, see Table
\ref{spectrap2}). Thermal emission from a small polar region, with
very little modulation  and present in both radio modes, is
an obvious consequence of the pulsar geometry and orientation.
Fig. \ref{fig_area_kT} shows the values of projected emitting area  and temperature derived for the constant
blackbody component. The corresponding bolometric luminosity is comprised between $\sim$5$\times10^{28}$  and  $10^{29}$ erg s$^{-1}$.
The emitting area is smaller than that expected for a polar cap whose radius is determined
by  the last closed  magnetic field lines reaching the light cylinder in a dipolar geometry,
$R_{PC}$ = $R (2\pi R/Pc)^{1/2}$. For a neutron star radius $R$=10 km, this gives a polar cap area
$\sim\pi R_{PC}^2$ = 6$\times$10$^4$ m$^2$ for \src .
A similar situation is  observed in most old pulsars whose thermal emission is attributed to regions of the star surface heated by
back-flowing particles accelerated in the magnetosphere above the magnetic poles. A plausible scenario to account for the
small emitting area is that the X-rays come only from the footprints of the sub-beams which constitute the radio emitting
pattern of drifting radio pulses \citep{zha05}.

In the partially screened gap model \citep{gil03,gil06}, a simple relation is expected between the heating rate of the polar cap and
the drift velocity of the plasma sparks around the magnetic axis.
This depends on the ratio between the  circulation period, $P_4$, and the spin period.
The efficiency of the thermal emission is  $L_\mathrm X/\dot E=0.63\, (P/P_{4})^2$,  which for \psr\
($P_4=37.4 P$, \citealt{des99}) corresponds to a luminosity of 4.5$\times$10$^{28}$ erg
s$^{-1}$, in reasonable  agreement with the results of our fit.

The power-law component can be ascribed to non-thermal emission
produced in the magnetosphere with a typical efficiency of rotation-powered pulsars \citep{pos02}.
The average luminosity of \src\
(2$\times$10$^{29}$ erg s$^{-1}$,  2--10 keV) implies an
efficiency of 2$\times$10$^{-3}$ fully consistent with that of old
pulsars \citep{pos12}. The presence of pulsations in this component is a
natural consequence of a non-isotropic emission pattern.
A simple model in which the power-law component varies sinusoidally
while the blackbody is constant can reproduce the pulsed fractions
observed in the different bands quite well within the reported
uncertainties.  A slight variation in the beaming direction might
explain the difference between the B and Q mode, not necessarily
implying a real difference in the intrinsic luminosity.  This could be due to a minor change in the
structure of the magnetosphere and/or of the region where the relativistic particles responsible
for the non-thermal emission are accelerated.
It is also not surprising that such changes affect both the high-energy and the radio emission properties.

\section{Conclusions}

We have performed a detailed analysis of the 2011  \xmm\ observations of \src\ which confirms the striking X-ray
variability correlated with the radio mode  recently reported by \citet{her13}.
\src\ is a faint X-ray source: only $\sim$600 photons have been detected in 100 ks, despite the large
collecting area of the EPIC instrument. While there is no doubt that the X-ray flux of \src\ varies by
more than a factor two, being fainter when the radio emission shows a regular pattern of drifiting subpulses (B-mode),
all the other spectral and timing results are subject to uncertainties which leave open
different interpretations.

We advocate a scenario in which a baseline level of thermal  X-ray emission is always present and originates in a hot
polar region, visible at all rotational  phases and in both radio modes.
This is well fit by a   blackbody spectrum with temperature $kT$=0.26 keV and luminosity of $\sim$7$\times10^{28}$ erg s$^{-1}$,
in good agreement with the predictions of the partially screened  gap model.
The higher X-ray flux of the Q-mode can be explained by the appearance of an additional component,
modulated at the star spin period.
We found that this component can be described by a power-law with photon index 2.2, and
suggest that it is due to anisotropic non-thermal X-rays produced  in the neutron star magnetosphere.
A small change in the beaming pattern or in the efficiency of  acceleration of the particles responsible for the
non-thermal emission can  explain the reduced flux of this component during the radio B-mode,
without invoking global re-arrangements of the star magnetosphere.

No similar X-ray variability has been reported for other rotation powered neutron stars.
However it should be considered that, without an ad hoc analysis based on time intervals selected from
the simultaneous radio  data, the X-ray variations in \src\  would have passed unnoticed.
It is thus possible that other pulsars, in particular those alternating different
radio patterns and/or showing intermittent radio emission \citep{kra06},
can exhibit associated variations in their  X-ray properties which can only be unraveled by
sensitive X-ray observations with simultaneous radio coverage.

\section*{Acknowledgments}
This research is based on data of XMM-Newton, an ESA science mission with instruments and
contributions directly funded by ESA Member States and NASA.
This work was partially suported by the  PRIN-INAF 2010.

\bibliographystyle{mn2e}
\bibliography{pap_psrb0943}

\bsp

\label{lastpage}

\end{document}